# A Comparative Study with Traditional and Transfer Learning-enhanced Machine Learning Algorithms for Geotechnical Characterisation of Coal Spoil

Sureka Thiruchittampalam[a,b], Kuruparan Shanmugalingam[c], Bikram P Banarjee[d], Nancy F Glenn[e] and Simit Raval[a*]

[a] *School of Minerals and Energy Resources Engineering, University of New South Wales, Sydney, 2052, NSW;* [b] *Department of Earth Resources Engineering, University of Moratuwa, Moratuwa, 01400, Sri Lanka;* [c] *School of Computer Science and Engineering, University of New South Wales, Sydney, 2052, NSW, Australia;* [d] *School of Surveying and Built Environment, University of Southern Queensland, Toowoomba, 4350, Queensland, Australia;* [e] *Department of Geosciences, Boise State University, Boise, ID, USA*

\* simit@unsw.edu.au

# A Comparative Study with Traditional and Transfer Learning-enhanced Machine Learning Algorithms for Geotechnical Characterisation of Coal Spoil

The characterisation of materials is a prerequisite for evaluating and predicting the stability of mining waste dumps. Over the past three decades, the BHP Mitsubishi Alliance Coal framework has been a cornerstone in Australian coal mines for characterising waste dump materials. However, its reliance on subjective human observations has introduced potential inaccuracies and subjectivity into the process. In response to these limitations, this study proposes an innovative approach to classify coal spoil attributes by remotely acquiring images through phones/tablets. Automated image-based classification relies on feature extraction and a substantial amount of data. Nevertheless, the inherent complexity of geological factors contributing to the formation of both rare and dominant materials leads to imbalanced data. Recognising the need for classification mechanisms to overcome these challenges in spoil classification, the study explores and compares the use of convolutional neural networks, hybrid deep learning, and traditional techniques. Among the twelve models evaluated in this study, the ResNet18-k nearest neighbour model emerges as a powerful tool in geotechnical characterisation. However, it is essential to address issues of interpretability and adaptability to diverse datasets. As this study evolves, the field of geotechnical characterisation of spoil can anticipate the development of more robust methods in the future.

## 1  1 Introduction

The mining industry is experiencing a paradigm shift, driven by the compelling need to augment operational efficiency, mitigate risks, and optimise profitability. Central to this transformation is automated coal spoil characterisation, an indispensable component of contemporary mining practices. This evolution is pivotal in addressing a multitude of challenges, encompassing ensuring dump stability, facilitating actionable decision-making, and capitalising on the transformative potential of automation for sustainable mining practices (McQuillan, Canbulat, and Oh 2020; Ma et al. 2023). Delving into methodologies for real-time automated spoil classification holds promise for enhancing safety and laying the groundwork for the development of as-dumped strength models. These models can be seamlessly integrated for automated slope stability modelling, ushering in a new era of data-driven, dynamic stability assessment.

In the last two decades, the BHP Mitsubishi Alliance Coal (BMAC) framework has emerged as a prominent classification system for systematically characterising coal spoil based on visual-tactile attributes (Simmons and McManus 2004). Traditionally, this process heavily relies on descriptions of observed geological features, necessitating laborious and risky fieldwork and leaving room for subjective interpretation by experts (Bradfield, Simmons, and Fityus 2013). Application of deep learning, offers a data-driven approach (Phoon and Zhang 2023) by utilising observational data to create classification models, thereby enhancing the accuracy and objectivity of spoil classification.

Most real-world datasets exhibit a long-tailed distribution, wherein the training set showcases an exponential decay distribution in the sample size of each class (Dawson, Dubrule, and John 2023). Furthermore, the categorisation of coal spoil underscores an inherent imbalance, commonly referred to as the "imbalanced data problem". This class size imbalance is a consequence of geological variations within a site, arising from complex interactions among geological, tectonic, climatic, and environmental factors occurring over geological time scales. These interactions result in certain geological characteristics being rare while others become dominant. The core challenge lies in accurately predicting the minority classes within the dataset by leveraging the knowledge gained from the majority classes. Compared to traditional classification algorithms, deep neural networks show superior performance on class-balanced datasets (Dawson, Dubrule, and John 2023). However, they face challenges in imbalanced data classification, as they tend to prioritise learning from majority classes, with limited metrics for assessing minority class predictions. Additionally, the efficiency of implementing deep learning models on new datasets is affected by the complexity of the model, dataset compatibility, model adaptability, and the expertise of the implementer, often leading to a cumbersome process. To address these challenges, the introduction of a transfer learning model provides an effective solution for handling class size-imbalanced data within multi-class datasets.

The methodological framework underlies transfer learning consists of a series of sequential steps that collectively contribute to successfully adapting pre-trained models. The first crucial step involves selecting a pre-existing model that aligns with the intricacies and requirements of the target task. Following model selection, the pre-trained model is repurposed to suit the intended task. This repurposing can involve strategies such as using the entire model, specific segments, or even just the initial layers of the neural network. With their inherent ability to capture generalised features, the initial layers hold particular value for effective adaptation (Shatwell, Murray, and Barton 2023). As the neural network is further navigated, it becomes more specialised in the given task. Fine-tuning, a critical phase, entails adjusting the final layers of the neural network to align its learned features with the nuanced demands of the target task. This process imparts a tailored approach to the network's capabilities, enhancing its effectiveness in addressing the specific intricacies of the problem at hand. The incorporation of these stages within the transfer learning paradigm not only enables the efficient utilisation of pre-trained models but also substantially enhances both model performance and training efficiency. This approach tackles challenges stemming from limited data availability and the resource-intensive nature of training extensive models.

This study offers significant advancements in image-based classification techniques for geotechnical characterisation of spoil, providing the mining industry with enhanced tools for environmental monitoring and mine management decision-making. By enabling more accurate and efficient characterisation of mining environments, these refined techniques pave the way for optimised resource utilisation, minimised environmental impact, and enhanced operational efficiency.

## 2 Methodology

This study proposes a machine learning classification methodology tailored for the geotechnical characterisation of coal spoil. The methodology employs a combination of traditional machine learning models, convolutional neural network (CNN), and deep hybrid learning. The primary objective of this study is to classify coal spoil according to the BMAC framework attributes, utilising close-range images obtained in the field settings during mining operations. The input data for this investigation comprises coal spoil images, acquired through a mobile phone camera (iPhone 11), for the classification modelling of attribute categories. This innovative approach has the potential to revolutionise coal spoil classification, enhancing spoil management practices.

### 2.1  Background

Transfer learning is a widely adopted and effective strategy for training networks with limited datasets (Ullah et al. 2022). It involves pretraining a network on a substantial dataset like ImageNet, comprising 1.4 million images distributed across 1000 classes (Han, Liu, and Fan 2018). This pretrained network's adaptability across domains, utilising shared feature extraction, exemplifies deep learning's utility in data-scarce scenarios (Wu et al. 2018). Numerous pretrained models, such as AlexNet, ResNet, EfficientNet and Inception, are publicly available along with their learned parameters (Zhang, Liu, and Shi

2020). Practical implementation of pretrained networks includes fixed feature extraction and fine-tuning approaches (Swati et al. 2019). In this study, both approaches were employed.

The fixed feature extraction approach removes the fully connected layers of a pretrained network, while preserving its convolutional and pooling layers, known as the convolutional base. This convolutional base operates as an unchanging feature extractor (Rostami et al. 2023). Additional machine learning classifiers, such as random forests, or conventional fully connected layers, can be stacked on this fixed feature extractor. Consequently, the training process focuses exclusively on the added classifier, utilising a specific dataset of interest.

In contradistinction, the fine-tuning strategy encompasses the substitution of fully connected layers within the pre-trained model by a new set of fully connected layers tailored to the target dataset. It also encompasses the refinement of all or a subset of the kernels embedded within the pre-trained convolutional base via the application of backpropagation. This approach affords the discretion of fine-tuning all layers within the convolutional base or, alternatively, preserving the immutability of the selected earlier layers while affecting the fine-tuning process on the deeper strata (Li et al. 2020). This methodology draws impetus from the observation that early-layer features tend to manifest greater generality, encompassing attributes such as edges, which bear relevance across a diverse spectrum of datasets and task domains, while subsequently derived features progressively exhibit heightened specificity, attuning themselves to the particularities of a given dataset or task.

In this study of coal spoil attribute characterisation, both transfer learning approaches, specifically fixed feature extraction and fine-tuning, have been extensively investigated. In addition, it is essential to acknowledge that transfer learning imposes certain constraints on input data parameters such as image dimensions. Notably, while the height and width of an input image may exhibit arbitrary values, they cannot plummet to excessively diminutive proportions. As a mitigating measure, the introduction of a global pooling layer interposed between the convolutional base and the newly appended fully connected layers (Khademi, Ebrahimi, and Kordy 2022) commonly serves as an architectural refinement to address this constraint.

In conjunction with the advanced CNN techniques proposed herein, we also delve into the conventional image classification methodology (Bag of features) (Kumar et al. 2021), integrating an established machine learning classifier. This dual-pronged exploration facilitates a comprehensive evaluation of the proposed techniques' performance. Comparing CNN-based approaches with traditional approach for attribute detection yields insights into the detection process. The first approach in this study employs CNN with fully connected and softmax layers, while the second, deep hybrid learning, uses traditional machine learning classifiers in the classification head. These results are compared to those of a conventional technique.

A visual representation of the methodology forming the basis of the analytical pipeline is given in Figure 1.

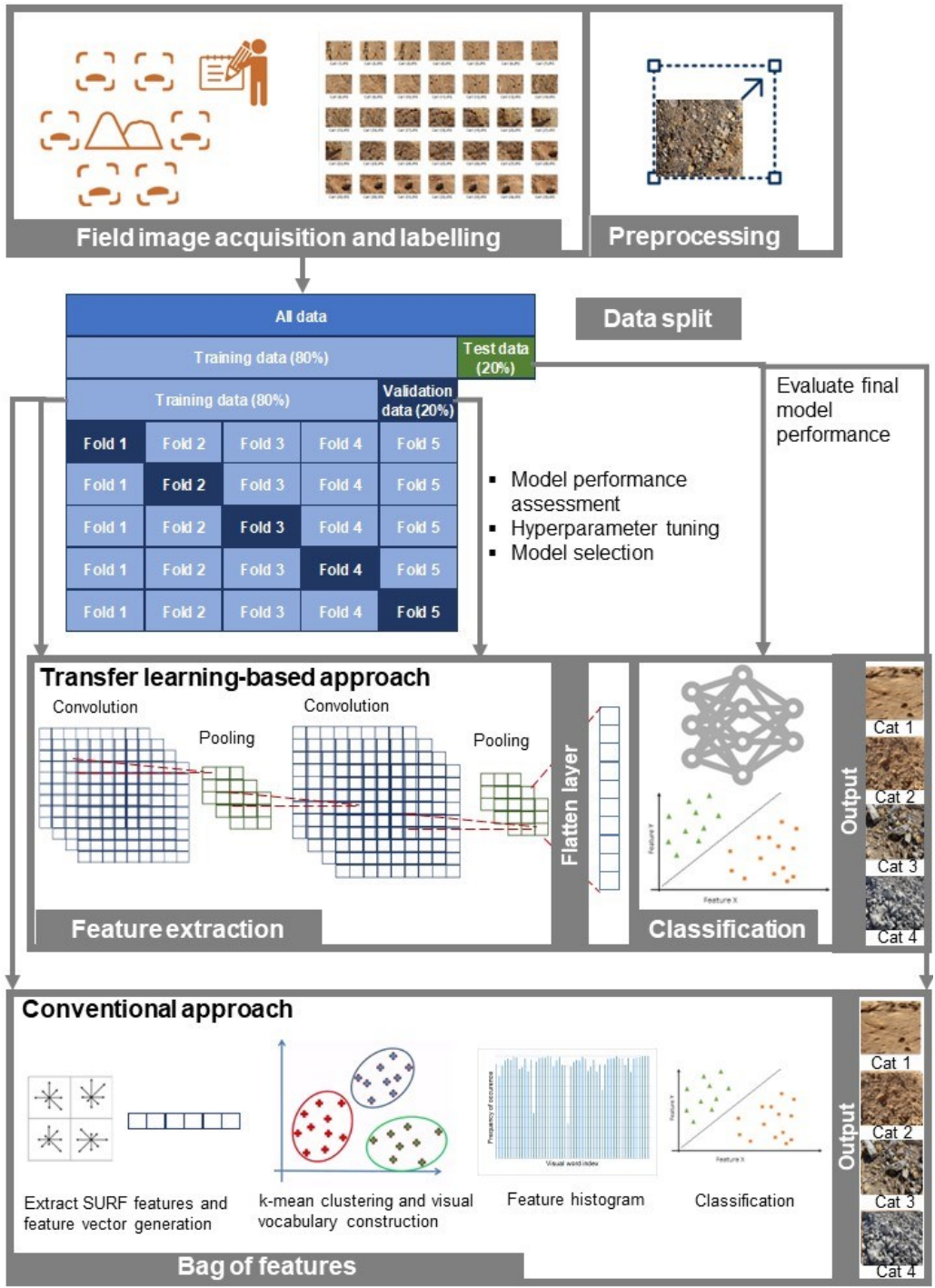

Figure 1. Schematic overview of the study's fundamental components. Image acquisition was conducted using handheld devices in field, the meticulous labelling of datasets was performed by spoil characterisation experts, ensuring the integrity of the data utilised in this study. Dataset partitioning employed stratified k-fold cross-validation (k = 5) to create training, validation, and test data subsets. The transfer learning-based approach consists of two main parts: the feature extraction component, represented by convolutional and pooling layers functioning as the convolutional base, responsible for extracting features, including edges, lines, and curves in lower layers, and shapes in upper layers, from the images. The classification component of the transfer learning approach includes fully connected and softmax layers (CNN) or traditional machine learning algorithms (deep hybrid learning), which facilitate image class predictions based on task-specific features. A Bag of Features approach was utilised in the same dataset to facilitate a comparative analysis with a conventional approach.

## 2.2  Data Collection

This study represents an essential initial step in the development of an image dataset dedicated to coal spoil within the mining sector, as no pre-existing dataset of this nature is available. The data acquisition process was initiated with the capture of images at a specific mine dump site located in New South Wales (NSW). These images depicted various categories of coal spoil available at the selected mine site. The information from the digital coal spoil images is represented by a 3D tensor with separate red, green, and blue (RGB) channels. Digital coal spoil images are 4032 pixels ×3024 pixels. A team of experts in the field categorised and annotated each image, allocating them to their respective categories based on BMAC framework (Table 1). Moreover, to augment the dataset's information, each image was furnished with a comprehensive set of attributes. The process of spoil characterisation involves consistently attributing specific qualities to the spoil and assigning weights to each of these attributes using standardised methods, as detailed in Table 1. "Predominant particle size" is determined through established visual techniques, while "consistency/ relative density" is assessed in the moist state, employing tactile procedures for cohesive or cohesionless materials. "Plasticity" can be determined either through a liquid limit test or visual examination. The "fabric structure" represents a spoil characteristic based on the distribution of particle sizes, consisting of two components: the "framework," which comprises larger particles transmitting forces and forming a rigid structure upon contact, and the "matrix," composed of smaller particles filling the gaps between framework particles. The BMAC categories are assigned by summing the relative weights of each attribute, and the spoil is then allocated to the category with the highest cumulative weight.

Table 1. Categories within the BHP Mitsubishi Alliance Coal (BMAC) framework for spoil characterisation, along with corresponding visual-tactile attributes (adapted from Simmons and McManus, 2004)

| Category → | Cat-1 | Cat-2 | Cat-3 | Cat-4 | Weightage ↓ |
|---|---|---|---|---|---|
| *Description* | *Fine-grained clay-rich high plasticity* | *Fine-grained low plasticity with larger clasts* | *Larger clasts with fine matrix, low plasticity* | *Large blocks, minor fines, minor slaking* | |
| Predominant particle size | Clay | Sand | Gravel | Cobbles | 11.6 |
| Consistency: (Cohesive) | Soft to Firm | Stiff | Hard | | 26.9 |
| Relative density: (Cohesionless) | Loose | Medium Dense | Dense | Hard packed | |
| Structure | 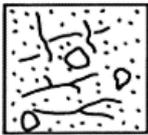 Matrix only | 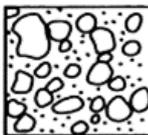 Matrix supported | 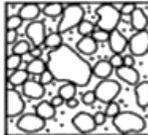 Framework supported | 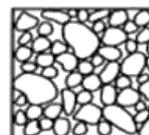 Framework only | 26.9 |
| Liquid Limit | High (>50) | Intermediate (35-50) | Low (20-35) | Not Plastic (<20) | 34.6 |

The framework, as outlined in Table 1, categorises spoil into one of four distinct categories and furnishes peak shear strength parameters, including cohesion ($c'$) and friction angle ($\phi'$), for three distinct strength mobilisation scenarios, namely unsaturated, saturated, and remoulded conditions (Table 2). This methodology facilitates a streamlined geotechnical characterisation of spoil, supplying essential empirical data for the scrutiny and substantiation of imaging-based analysis.

Table 2. BHP Mitsubishi Alliance Coal (BMAC) framework spoil categories and corresponding shear strength parameters for different mobilisation modes (after Simmons and McManus, 2004)

| Category → | | Cat-1 | Cat-2 | Cat-3 | Cat-4 |
|---|---|---|---|---|---|
| Unsaturated | $\gamma$ (kN/m$^3$) | 18 (1) | 18 (1) | 18 (1) | 18 (1) |
| | c' (kPa) | 20 (10) | 30 (15) | 50 (15) | 50 (15) |
| | $\phi'$ (deg) | 25 (2.5) | 28 (3) | 30 (2) | 35 (2.5) |
| Saturated | $\gamma$ (kN/m$^3$) | 20 (1) | 20 (1) | 20 (1) | 20 (1) |
| | c' (kPa) | 0 (0) | 15 (7.5) | 20 (10) | 0 (0) |
| | $\phi'$ (deg) | 18 (3) | 23 (2.5) | 25 (2.5) | 30 (1.5) |
| Remoulded | $\phi'$ (deg) c' = 0 kPa | 18 (1.5) | 18 (1.5) | 18 (1.5) | 28 (2) |

The characterisation of coal spoil is grounded in the study conducted by Simmons and McManus (2004). Their study led to the development of the BMAC spoil shear strength framework, as outlined in tables 1 and 2. This framework has gained widespread acceptance as a means of categorising coal mine spoil. It relies on visual and tactile properties to evaluate the shear strength parameters of the spoil, thus obviating the need for time-consuming laboratory tests.

To address in-field characterisation limitations, this study utilised curated images for automated characterisation. The collected 2095 image dataset encompasses multiple categories (Cat-1 to Cat-4) along with their corresponding attributes (Figure 2). The selected site features a limited set of categories within the BMAC classification. Figure 2 offers an overview of the available categories on the specified site, along with the corresponding proportions for each category.

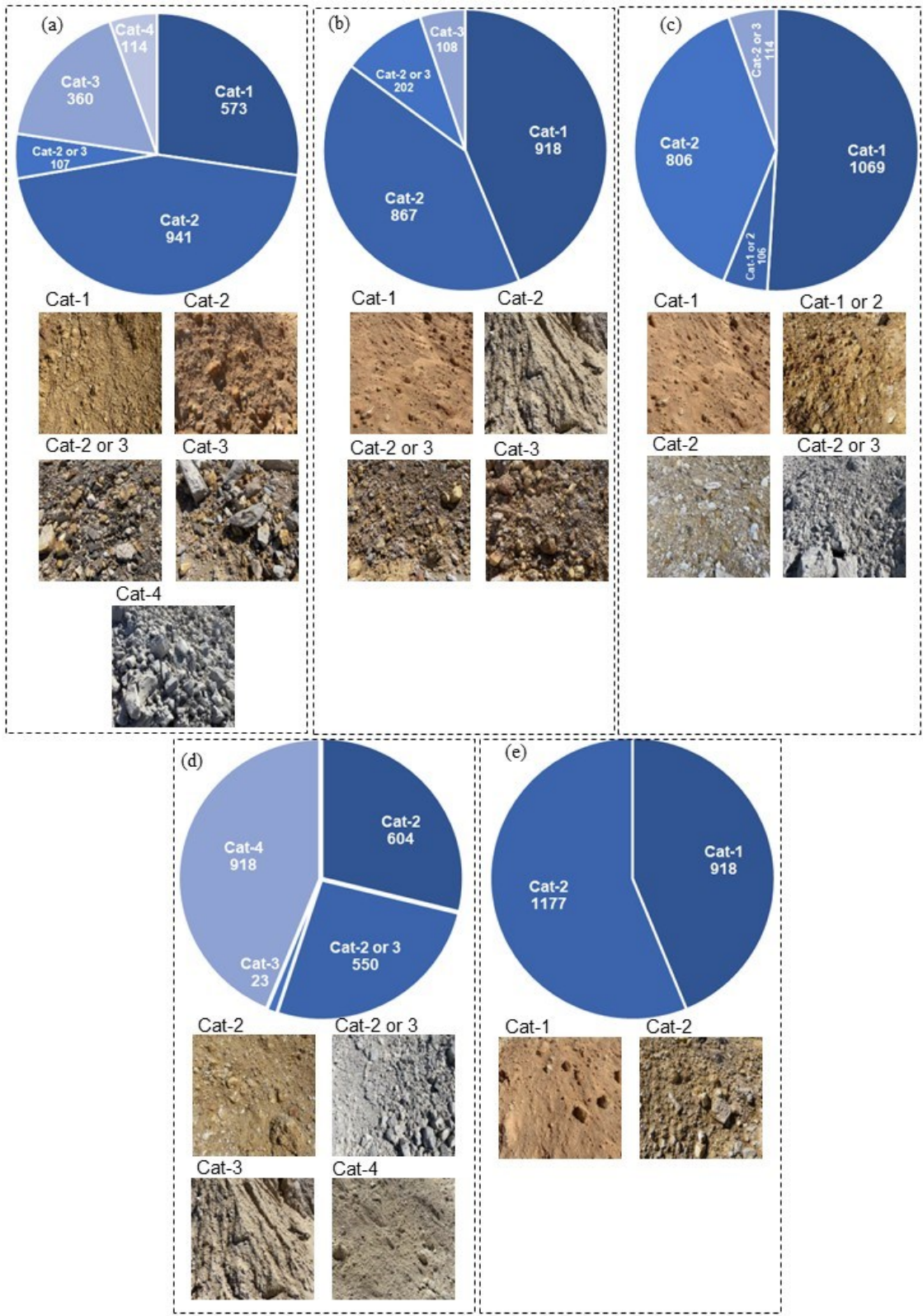

Figure 2. Distribution of spoil images for each category in (a) particle size distribution, (b) relative density, (c) fabric structure, (d) plasticity and (e) BMAC category.

*2.3 Data preparation*

Preceding algorithm deployment, resizing all dataset images to a uniform dimension is imperative, especially for many deep learning algorithms. For the dataset to be compatible with the original dimensions of the pre-trained architectures, and to leverage the natural-image features learned by the pretrained networks, the images used in this work were resized to 512 × 512 pixels.

Ensuring the robustness of predictions represents a fundamental consideration when evaluating the machine learning algorithms. In the context of this investigation, this issue was systematically addressed through the partitioning of each dataset, allocating 80% for training data and reserving a 20% segment for test data, thus establishing a rigorous evaluation framework. Furthermore, the training data was subjected to cross-validation, resulting in the establishment of 5 distinct folds, encompassing 80% for training purposes and 20% for validation purposes. This validation subset accounted for 16% of the entire dataset. Given the substantial class size imbalance present within the datasets, we opted for a stratified k-fold cross-validation approach, utilising k equals 5. This method kept the balance of samples in each class consistent across all folds (Zhang et al. 2019), ensuring that the class distribution remained the same as in the entire training dataset. By doing so, the susceptibility of the model's performance to the specific randomisation of data partitions is minimised. Consequently, this approach enhances the robustness of the performance evaluation, rendering it more representative of the model's capacity for generalisation across diverse datasets.

*2.4 Model training*

*2.4.1 Transfer learning-based approach*

In this study, we employed a range of pre-trained convolutional networks (Table 3), including AlexNet, ResNet18, ResNet50, InceptionV3, and EfficientNetB0. AlexNet, with eight layers, is an early CNN, while ResNet18 and ResNet50, being deeper (18 and 50 layers respectively), use residual connections. InceptionV3 employs inception modules for multiscale features, and EfficientNetB0 prioritises efficiency and performance balance (Elhassouny and Smarandache 2019; Yang et al. 2021).

Fixed feature extraction and fine-tuning strategies were applied to adapt these pre-trained models to the specific attribute classification task at hand. This approach capitalises on the pretrained models, tailoring them to the finer distinctions required for attribute classification, thereby enhancing the models' performance and efficiency.

Table 3. Overview of the models used in this comparative analysis.

| Model | Total learnable | Layers | Learnable layers |
|---|---|---|---|
| AlexNet | 58.5M | 25 | 8 |
| ResNet18 | 11.1M | 71 | 18 |
| ResNet50 | 23.5M | 177 | 50 |
| InceptionV3 | 21.8M | 315 | 48 |
| EfficientNetB0 | 4M | 290 | 82 |

The employment of these networks' pre-trained weights and parameters facilitated the extraction of generic features from the spoil pile images. Moreover, we modified the network's final classification layer, replacing it with a new classifier that was trained specifically on the spoil image dataset. This new classification head was implemented by substituting the last three layers with a fully connected layer, a softmax layer, and a classification output layer. Additionally, we conducted training experiments starting from zero weights to train the models from scratch. In certain cases, the new classification head was replaced by alternative classifiers, such as k-nearest neighbours (knn), decision tree (DT), support vector machine (SVM), and ensemble methods (Figure 3).

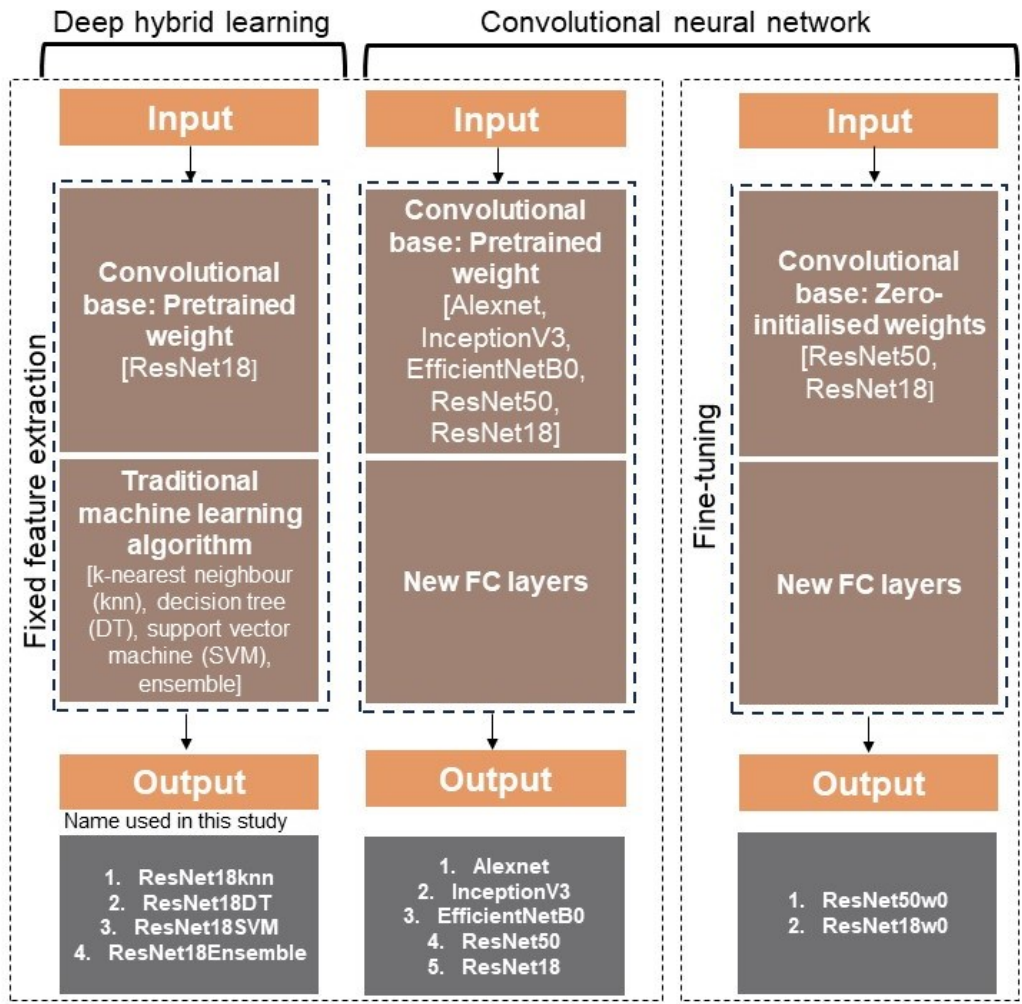

Figure 3. Employed models with their weights and classification heads.

2.4.1.1   Convolutional neural network (CNN)

Model training of CNN involves the iterative adjustment of weights and biases within the convolutional and fully connected layers' kernels, aiming to minimise disparities between predicted and true labels within the training dataset. The assessment of network performance occurs through a cost function, which gauges the divergence between output predictions and actual labels via forward-propagation. Given the nature of multi-class classification, the cross-entropy loss was adopted as our chosen cost function (Bokhorst et al. 2023). Subsequently, the softmax function was applied to the cross-entropy outputs, thereby assigning a class to each sample. The refinement of learnable parameters transpires iteratively, guided by the loss value, as informed by backpropagation and gradient descent optimisation techniques (Yaqub et al. 2020).

To achieve this, the Adam gradient descent algorithm, a widely embraced optimisation technique in deep learning, was employed. A constant learning rate of $10^{-5}$ was maintained throughout the experiments. It is noteworthy that the Adam algorithm integrates adaptive learning rates, amalgamating elements of both momentum and RMSprop (Wang et al. 2020). This adaptive learning rate mechanism facilitated efficient parameter updates during the training process, thereby contributing to the model's convergence. A uniform batch size of 64 was employed across all models.

The final model selection was based on the convergence in both accuracy and loss curves. Upon convergence, the model that had undergone training was chosen for the accuracy of the test data. To mitigate random fluctuations in CNN performance, training was conducted five times. The performance of each of the five models generated during these iterations was individually employed to make predictions on the test data.

#### 2.4.1.2 Deep hybrid learning

Deep hybrid learning omits FC layers and integrates a machine learning algorithm, such as SVM classifier, into the feature extractor. The foundational structure of the provided networks remains consistent in other variants. To fortify the classifier's robustness, the training process underwent five iterations. Each of the five models produced during these iterations was utilised independently to generate predictions on the test data.

#### *2.4.2 Traditional machine learning approach*

The conventional approach employed in this study was based on the bag of features. This involved a series of steps, starting with the extraction of speeded-up robust features (SURF) from all images within various image categories. The SURF algorithm is employed to compute descriptors for the points of interest, which form the basis for subsequent image analysis.

The construction of the visual vocabulary was the next element of this approach. To create this vocabulary, the feature space was quantised using k-means clustering. This process effectively reduced the dimensionality of the features and ensured that each feature could be associated with a specific visual word, which was pivotal in image representation and classification.

An approximate nearest neighbour algorithm was employed to facilitate efficient feature matching with the visual words in the constructed vocabulary. This step streamlined the processing of image features and improved the computational efficiency of the method (Li, Wang, and Zhang 2016). Following feature matching, histograms were generated for each image, calculating the occurrences of visual words. These histograms represented a novel and condensed representation of the images, encoding their distinctive visual characteristics. Importantly, the length of the histogram corresponded to the number of visual words, ensuring consistency in image representation.

To create the training data, the procedure was applied to the images within the training set. These histograms formed the basis for training a classifier (Srivastava, Bakthula, and Agarwal 2019). In this study, an SVM classifier was chosen for its robust classification capabilities.

To enhance the robustness of the classifier, the training process was iterated five times. Each of the five models generated during these iterations was utilised independently to make predictions on the test data. This comprehensive methodological framework, underpinned by the bag of feature principles, SURF features, k-means clustering, and SVM classification with linear kernel function, provided an effective approach for spoil image categorisation.

## 2.5 *Evaluation*

Subsequent to the training phase, the assessment of predictive capabilities for distinct models on the coal spoil test dataset involved the utilisation of several evaluation metrics. Notably, overall accuracy served as a fundamental metric, quantifying the ratio of accurately predicted classifications to the total count of classified images.

$$Overall\ accuracy = \frac{TN + TP}{TN + TP + FN + FP}$$

In this context, true negative (TN) represents the count of instances accurately identified as negatives, true positive (TP) signifies the count of instances correctly identified as positives, false negative (FN) denotes the count of instances erroneously classified as negatives, and false positive (FP) indicates the count of instances erroneously classified as positives. The metric of overall accuracy, in its uniform treatment of correctly predicted classifications across various classes without differentiation, may be regarded as a potentially misleading measure. This concern is particularly pertinent in situations characterised by substantial class size imbalances (Fig. 2), where classes with a higher number of samples can exert a disproportionate influence on this statistic. Hence, we employed multiclass averaging as a method to calculate the mean per class accuracy (MPCA) (Dawson, Dubrule, and John 2023). This MPCA is defined as the unweighted mean obtained by summing the accuracies for each distinct class independently.

$$MPCA = \frac{\sum Overall\ accuracy\ of\ each\ class}{Number\ of\ classes}$$

Precision quantifies the ratio of correct positive identifications, indicating the probability that the predicted class for the spoil category truly belongs to that particular class.

$$Precision = \frac{TP}{TP + FP}$$

Recall determines the ratio of true positives among all actual positives that were accurately identified.

$$Recall = \frac{TP}{TP + FN}$$

The performance of all five iterative models generated during the five-fold cross-validation on the test data was assessed using the respective evaluation metrics. Following this, the mean and standard deviation of these evaluation metrics were calculated. Ultimately, comparisons between models were conducted using the mean values of the evaluation metrics.

Further, to determine if the best-performing model exhibited a significantly better performance than the other models, a two-tailed t-test was conducted at a significance level of 5%.

## 3 Results

### *3.1 Performance evaluation for overall accuracy*

#### *3.1.1 Convergence analysis of training and validation accuracy and loss curves in CNN*

The loss function evaluates a model's performance in a dataset, indicating its effectiveness after each optimisation epoch. In training CNN, the goal is to minimise errors computed by the loss function while improving testing accuracy. The ResNet50 model, applied to coal spoil attributes, demonstrates superior performance compared to other CNN models. In Figure 4, the convergence of accuracy and loss curves during training for the best-performing fold is illustrated. In this optimal fold, ResNet50 achieved the highest validation accuracy and the lowest validation loss across various attributes, namely particle size distribution (96.61%, 0.1158), relative density (95.11%, 0.1483), fabric structure (98.21%, 0.06278), plasticity (97.02%, 0.08776), and BMAC category (99.22%, 0.02732). These outcomes emphasise the effectiveness of the ResNet50 model in minimising loss and optimising accuracy for the classification of coal spoil attributes, surpassing all other CNN models employed in this study.

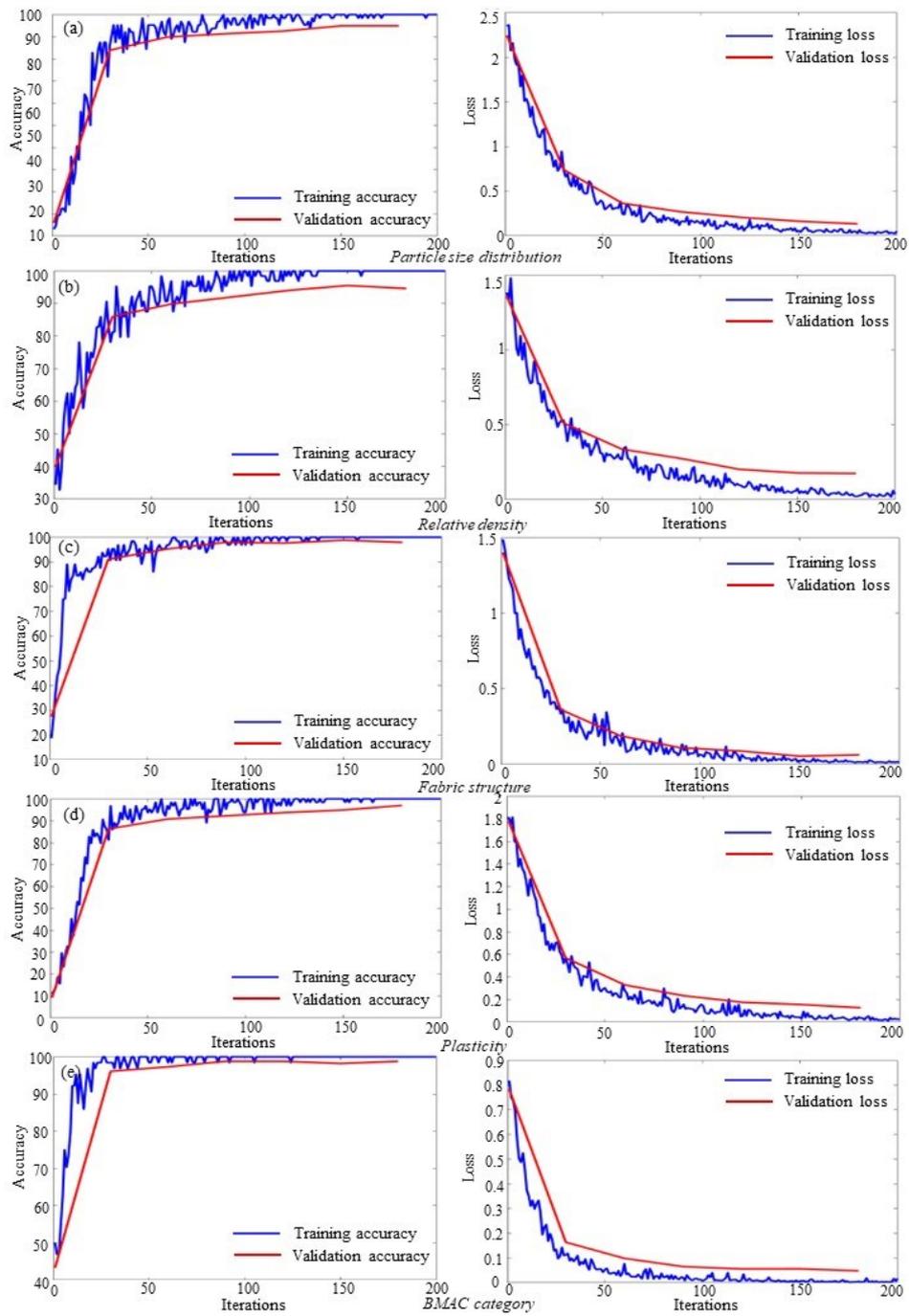

Figure 4. Learning curves illustrating the training and validation accuracy and loss obtained using the pre-trained ResNet50 network.

*3.1.2 Comparison of overall accuracies of traditional and transfer learning-based (CNN and deep hybrid learning) machine learning approaches*

ResNet18knn consistently achieves superior accuracy across all attributes (Figure 5), including particle size distribution, fabric structure, relative density, plasticity, and BMAC category, with respective mean accuracies and standard deviation of 99.28 (±0.35)%, 99.92 (±0.10)%, 99.08 (±0.39)%, 99.78 (±0.16)%, and 99.88 (±0.10)%. In contrast, the bag of features model consistently demonstrates comparatively inferior performance, registering accuracies of 80.04 (±1.22)%, 44.28 (±6.59)%, 34.4 (±12.61)%, and 66.88 (±3.46)%, respectively, for the mentioned categories, excluding BMAC category. ResNet18Ensemble records the lowest accuracy of 53.94 (±2.19)% for the BMAC category. The performance gap highlights the importance of selecting models for specific categories. ResNet18knn's commendable performance, despite modest computational demands, underscores the significance of making optimal model choices in classification tasks.

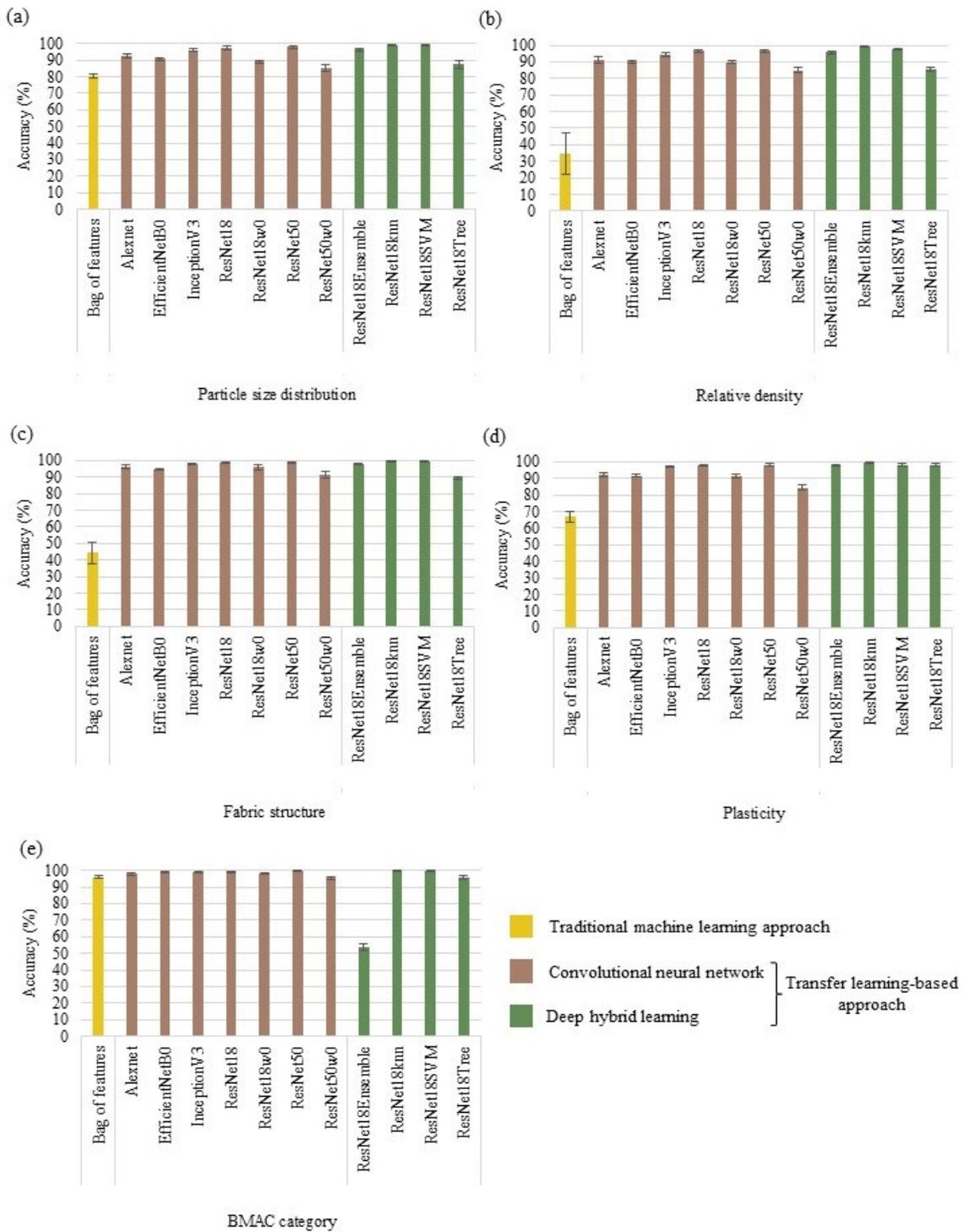

Figure 5. Comparative performance evaluation of classification accuracy in the test dataset for (a) particle size distribution, (b) relative density, (c) fabric structure, (d) plasticity, and (e) BMAC category.

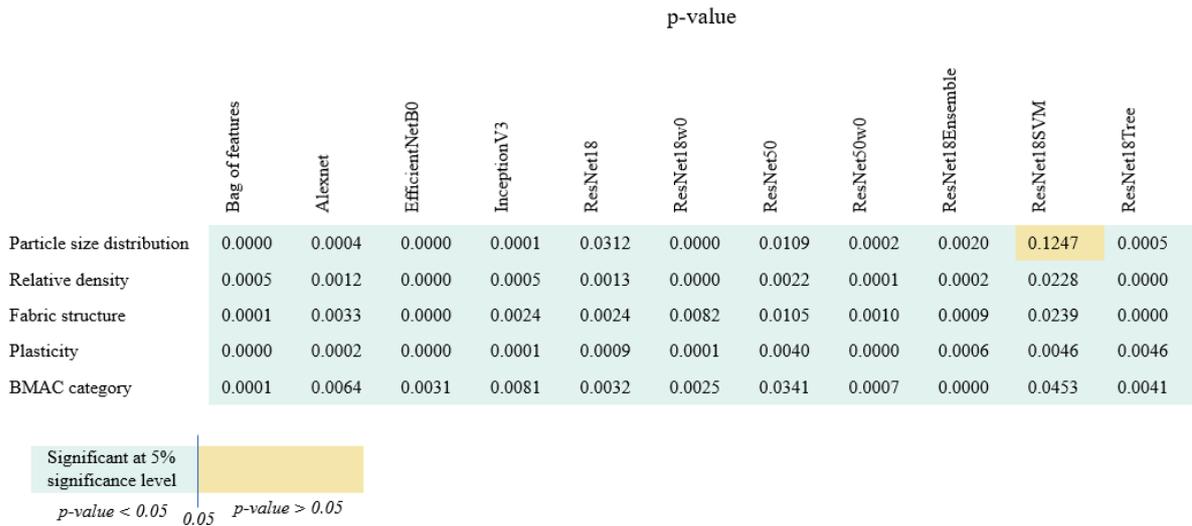

Figure 6. p-values from t-test evaluating the significance of performance differences between various models and the best-performing model, ResNet18knn.

According to Figure 6, with the exception of ResNet18SVM in the context of particle size distribution classification, all other models exhibit a statistically significant difference in performance compared to ResNet18knn across all attributes at 5% significance level. Moreover, ResNet18knn consistently demonstrates a higher mean overall accuracy than all other models, underscoring its significant superiority in performance.

### 3.2 Performance evaluation for each class

The performance metrics for the best performing classification model, ResNet18knn, across categories within attributes, are presented in Figure 7.

Particle size distribution, relative density, fabric structure, plasticity and BMAC category achieves MPCA of 99.71 (±0.14)%, 99.55 (±0.19)%, 99.95 (±0.06)%, 99.88 (±0.08)% and 99.90 (±0.12)%, affirming reliable instance categorisation (Fig. 7(f)).

ResNet18knn consistently achieves high precision and recall across all the attributes (Fig. 7(a)-(e)): particle size distribution (precision > 0.91, recall > 0.98), relative density (precision > 0.93, recall > 0.95), and fabric structure, plasticity, and BMAC category

(precision and recall > 0.99). Across attributes, distinct patterns emerge regarding precision and recall metrics. In particle size distribution, Cat-2 or 3 exhibits the lowest precision, while Cat-1 has the lowest recall. In relative density classification, Cat-3 demonstrates the lowest precision, and Cat-2 or 3 shares the lowest recall. For fabric structure, Cat-1 attains the lowest precision, whereas Cat-2 records the lowest recall. Plasticity classification reveals Cat-2 with the lowest precision, and Cat-2 or 3 with the lowest recall. Lastly, in BMAC category classification, both the categories showed precision and recall values above 0.90. These high values suggest that ResNet18knn is effective at minimising both false positives and false negatives, making it a promising model for spoil classification.

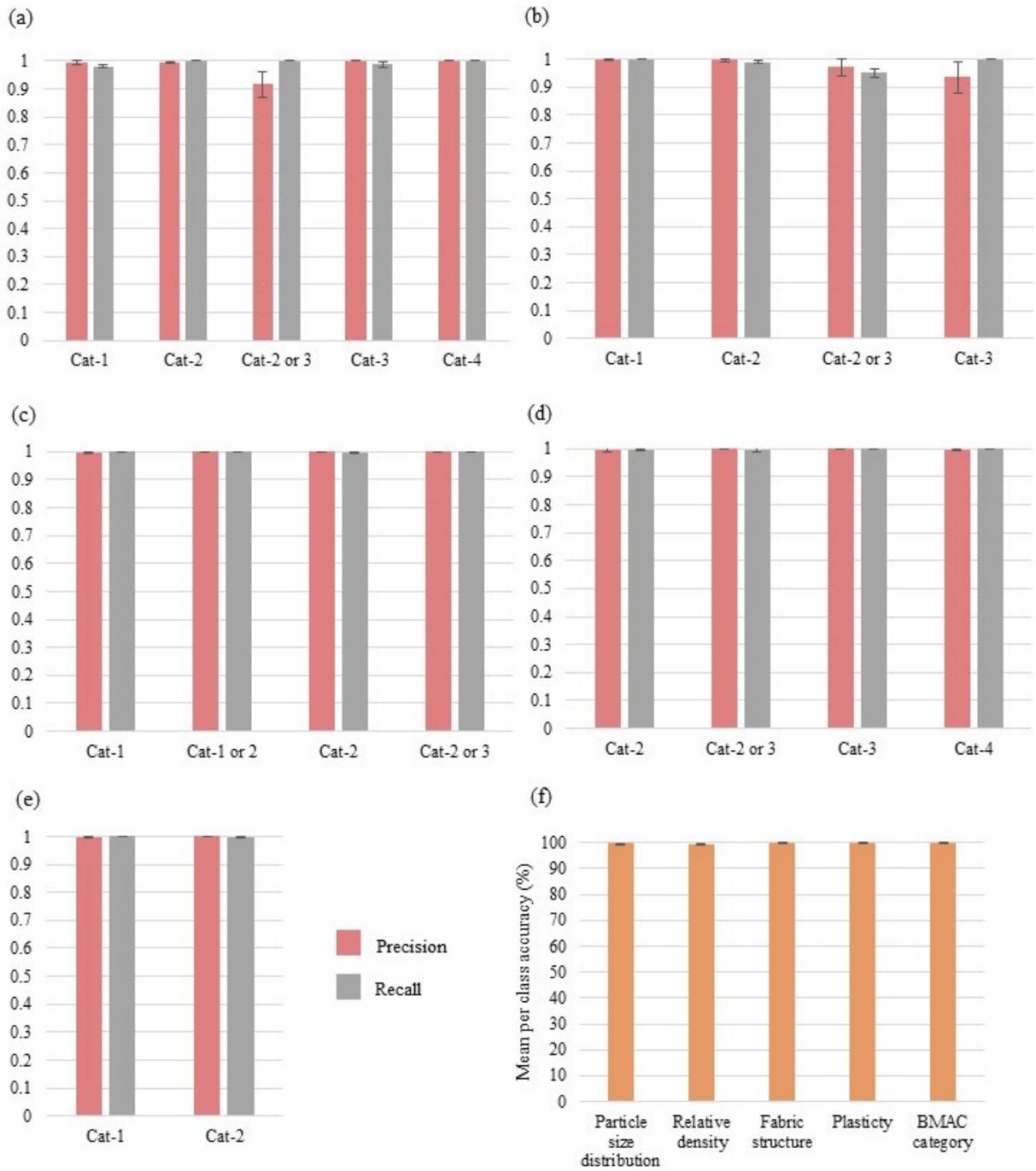

Figure 7. Precision and recall on dataset for (a) particle size distribution, (b) relative density, (c) fabric structure, (d) plasticity, and (e) BMAC category for Resnet18knn. (f) Mean per class accuracy for all attributes when deploying ResNet18knn.

## 4   Discussion

### *4.1   Comparison of transfer learning-based machine learning models*

A comparative analysis within each of the two primary categories of models in transfer learning-based machine learning, specifically CNN and deep hybrid learning models, provides additional insights into the performance of models within these respective categories.

#### *4.1.1   Comparison of convolutional neural networks*

The assessment of model complexity is crucial to mitigate the potential risk of overfitting and reduce computational resource demands. This evaluation involves a comprehensive analysis, considering factors such as the total number of learnable parameters, the number of layers, and the number of learnable layers (Hussain, Bird, and Faria 2019). In terms of number of learnable layers within a network, EfficientNetB0, for instance, stands out by having the highest number of learnable layers among the models considered. This characteristic indicates that a substantial portion of its layers actively participates in weight updates during training, potentially enhancing its capacity for feature extraction and representation learning. Nonetheless, it is imperative to acknowledge that the relationship between model complexity and performance is not unilaterally positive. More intricate models, despite their superior capacity for pattern recognition, may be prone to overfitting when applied to smaller datasets (Horry et al. 2020). Furthermore, they often necessitate significantly greater computational resources for training and inference. Consequently, the decision regarding the appropriate level of model complexity should be contingent upon the specific task at hand and the available computational resources, ensuring an optimal balance between model capability and practical feasibility.

In this study, the dataset under examination encompasses a total of 2095 images, serving as the foundational framework for exploration into the effectiveness of various deep learning models in spoil characterisation. Within this experimental framework, focusing exclusively on deep CNN models, models are hierarchically organised from deeper to shallower networks, as follows: EfficientNetB0, ResNet50, InceptionV3, ResNet18, and Alexnet. Among these, it is ResNet50 that emerges as the standout performer, exhibiting superior performance when compared to its counterparts. ResNet50 distinguishes itself by its innovative approach to mitigating the vanishing gradient problem (Reddy and Juliet 2019). This problem, often encountered in deep neural networks, hampers training by impeding the effective flow of gradient information through layers. To address this issue, ResNet50 introduces residual blocks, incorporating skip connections between layers. This strategic design choice facilitates a smoother and more efficient flow of information and gradients through the network, thereby enhancing training and convergence (Wen, Li, and Gao 2020). Hence, the remarkable performance of ResNet50 in spoil image classification tasks can be attributed to a combination of factors, including its architectural innovation, and its effective mitigation of the vanishing gradient problem. The network's ability to capture and represent intricate patterns and features within images is another factor in its success, as it enables the precise discrimination of spoil-related characteristics within the dataset.

It is also worth noting that within the ResNet50 family, the model fine-tuned from scratch, designated as ResNet50w0, exhibits the least favourable performance among our examined deep CNN models. The relatively suboptimal performance of ResNet50w0 can be attributed to the nature of training a deep neural network from scratch (Sahoo et al. 2022). Fine-tuning a model without leveraging pre-trained weights requires a substantially larger annotated dataset and prolonged training time to achieve an acceptable level of performance. In contrast, the pre-trained ResNet50 model, which leverages knowledge from a broad dataset, has a significant advantage in both initialisation and convergence speed.

In summary, the superior performance of ResNet50 within our study can be attributed to its unique architecture, specifically designed to mitigate the vanishing gradient problem and effectively process image data.

*4.1.2  Comparison of deep hybrid learning models*

When comparing deep hybrid learning models (ResNet18 with Ensemble, SVM, DT and knn), ResNet18knn emerges as the standout performer among the models and other transfer learning model-based approaches. Several factors contribute to the superior performance of ResNet18knn. Firstly, its architecture, based on ResNet-18, is known for its effectiveness in image classification tasks due to its deep architecture with residual connections. Additionally, the incorporation of a knn component in ResNet18knn allows it to capture local patterns effectively, which is particularly beneficial for this specific task. This study underscores the significance of empirical evaluation when selecting a model for spoil characterisation, emphasising that model performance can be influenced by numerous factors, including architecture, classifiers, data preprocessing, and domain-specific insights.

The results highlight the exceptional performance of the ResNet18knn model across various datasets and categories, showcasing its proficiency in spoil classification tasks.

## *4.2  Comparison of traditional machine learning models with transfer learning-based machine learning models*

Bag of features performed low when classifying particle size distribution, relative density, fabric structure and plasticity. Traditional machine learning algorithms may struggle with limited data and complex feature extraction (Sharma and Mehra 2020), while transfer learning allows the use of pre-trained models that have learned relevant features from large datasets. Transfer learning along with traditional machine learning offers the benefits of leveraging pre-trained models, addressing data scarcity, and combining the strengths of both approaches for improved performance in spoil classification tasks. Thus, ResNet18knn capitalised on the benefits of feature extraction from ResNet18 and the advantages of knn. Notably, ResNet18knn demonstrated statistically significant superior performance compared to all other models. knn is a non-parametric, instance-based learning method that relies on the similarity between data points. It can perform well when the data exhibits clear clustering patterns or when the decision boundaries are highly non-linear. If the data is naturally clustered in a way that is conducive to knn's nearest-neighbour approach, it may outperform other methods. Since the dataset is imbalanced (i.e., one class has significantly fewer samples than others), knn may perform relatively well, as it does not make strong assumptions about class distribution. SVM, decision trees, or ensemble methods may require more specialised techniques or resampling strategies to handle imbalanced data effectively.

### *4.3   Performance evaluation of each class with best model*

The model consistently achieves precision and recall values above 0.90 for most categories across all datasets. This is a remarkable achievement, as high precision ensures accurate positive predictions, while high recall indicates the model's ability to capture all relevant positive instances. These results suggest that the ResNet18knn model excels in both making precise positive predictions and minimising false negatives, a crucial balance in classification tasks. The mean per class accuracy, which averages the model's performance across all categories, consistently attains near-perfect scores (>99%). This demonstrates the model's ability to maintain high precision and recall levels across a diverse set of categories within each dataset.

The ResNet18knn model's exceptional performance is dataset-specific, with perfect scores across all key metrics in each dataset. This highlights the model's adaptability and robustness in handling classification of spoil attributes.

The analysis further emphasises misclassification between specific attribute categories, such as Cat-2 and Cat-2 or 3, in relative density. One plausible explanation for this bias is that it could have been introduced during the human labelling of the spoil images. In future work, our goal is to mitigate interpreter bias by creating a dataset that includes multiple classifications made by different individuals. Additionally, it is conceivable that the CNN architecture itself may exhibit a bias toward salient features. To address this concern, the inclusion of more labelled images depicting borderline cases in the dataset could likely lead to improved accuracy. Furthermore, involving multiple individuals in the classification of these images would help highlight any discrepancies.

While the ResNet18knn model has showcased exceptional performance, there are areas for further research and improvement. One avenue of exploration is the model's generalisation across datasets with varying characteristics. Investigating the transferability of the model's knowledge to new and unseen datasets could be beneficial. Additionally, understanding the model's interpretability and potential biases in its predictions is vital for transparent deployment in spoil characterisation.

## 5    Conclusion

This study has explored the nuances of evaluating the performance and complexity of models within spoil characterisation, using a combination of CNN, deep hybrid learning, and traditional machine learning approaches. It highlights the delicate balance between model complexity and effectiveness, emphasising the need to tailor model selection to specific tasks while considering computational feasibility.

The comparison between transfer learning-based models and traditional machine learning algorithms revealed the statistically significant superiority of ResNet18knn, underscoring the influence of architecture and classifier selection on spoil characterisation tasks. The ResNet18knn model's performance, characterised by remarkable precision, recall, and mean per-class accuracy, establishes it as a robust and versatile choice for spoil attribute classification tasks.

Potential areas for future research include delving into the model's ability to generalise across diverse spoil image datasets, augmenting its interpretability, and mitigating potential biases within its predictions. This study not only furthers the understanding of model intricacies but also establishes a foundation for practical applications in spoil classification. The ability to accurately classify spoil attributes can streamline spoil handling processes, optimise resource allocation, and minimise environmental impacts. The ResNet18knn model's effectiveness in this regard offers a promising solution for enhancing mining operations and environmental sustainability.